\documentstyle[12pt]{article}

\def\frac#1#2{{\textstyle {#1 \over #2}}}

\def\Eq{\begin{equation}}   \def\Endeq#1{\label{#1} \end{equation}}
\def\Eqa{\begin{eqnarray}}  \def\Endeqa#1{\label{#1} \end{eqnarray}}

\def\O{ {\cal O} }

\def\VN{ V_{\nu} (r) }

\begin{document}
\begin{titlepage}

\begin{center}
September, 1992      \hfill       HUTP-92-A041\\
                  \hfill       UFIFT-HEP-92-28\\
\vskip .5 in
{\large \bf Long Range Forces from Two Neutrino Exchange Revisited}
\vskip .3 in
{
  {\bf Stephen D.H. Hsu}\footnote{Junior Fellow, Harvard Society of
     Fellows. Email: \tt Hsu@HUHEPL.bitnet, Hsu@HSUNEXT.Harvard.edu}
   \vskip 0.3 cm
   {\it Lyman Laboratory of Physics,
        Harvard University,
        Cambridge, MA 02138}\\
   \vskip .3 in

  {\bf Pierre Sikivie}
   \vskip 0.3 cm
   {\it  Department of Physics,
        University of Florida,
        Gainesville, Fl 32611}\\ }

  \vskip 0.3 cm
\end{center}

\vskip .5 in
\begin{abstract}
The exchange of two massless neutrinos
gives rise to a long range force which couples to weakly charged
matter. As has been noted previously in the literature, the potential
for this force is $\VN \propto G_{F}^2 / r^5$ with
monopole-monople, spin-spin and more complicated interactions.
Unfortunately, this is far too small to be observed in present day
experiments. We calculate $\VN$ explicitly in the electroweak theory,
and show that under very general assumptions forces arising
from the exchange of two massless fermions can at best yield
$1 / r^5$ potentials.

\end{abstract}
\vskip 3 in

\end{titlepage}

\renewcommand{\thepage}{\arabic{page}}
\setcounter{page}{1}

The prospect of discovering a new long range force coupling to
ordinary
matter is exciting from both the theoretical and experimental points
of view
\cite{longrange}. Since long range forces require the existence of a
massless particle, a logical place to look in the electroweak theory
is at
the effect due to neutrinos. The exchange of a single neutrino (or in
general
a single fermion) cannot
give rise to a force since the interaction changes the angular
momentum of
the sources involved. However, the exchange of two neutrinos can
leave the
quantum numbers of the sources unchanged, and hence can lead to a
long
range force.

One might guess on the basis of dimensional analysis that
the potential for this interaction could take the form
$\VN \sim  G_F^2 m^2 / r^3$,
where $m$ is the mass of the source particle. Feynman
considered this form when contemplating neutrinos as the mediators of
a
gravity-like interaction \cite{Feynman}.
If this were the interaction,
the effects of such a force might be observable in laboratory tests
\cite{lab}. At $r \sim cm$, the resulting force on normal
matter would be
roughly $10^{-6}$ times smaller than that due to gravity if $m$ is
the electron
mass, and comparable to that of gravity if $m$ is a nucleon mass.
The current limit
on deviations from a $1/r$ potential at $r \sim cm$
are of order $10^{-4}$ \cite{lab}.
If the two neutrino force were measurable, it would provide
experimental
information on neutrino masses which is complementary to that
obtained from
standard particle physics experiments - long range force experiments
are
sensitive to extremely small masses.

Unfortunately, the form given above for $\VN$ is incorrect.
The correct behaviour, which we will derive below, is $1/r^5$. This
yields a much smaller effect. The exact form of the interaction is
therefore
somewhat academic, but seems to us worth computing. The two neutrino
force was investigated previously by G. Feinberg and J. Sucher
\cite{FS} and by
A. De Rujula, H. Georgi and S. Glashow (unpublished).
In fact, almost all of the results which appear in this paper have
been obtained earlier by Feinberg and Sucher.
However, our method of computation is different and we feel that
it is sufficiently simple to warrant exposition.
We became aware of the earlier work only after completing our own
calculations.
The previous authors come to conclusions similar to ours, but our
detailed results disagree slightly with those of Feinberg and Sucher.
Our
potential $\VN$ is smaller than
theirs by a factor of two, and the coefficient of our $\sigma_1 \cdot
\sigma_2$ term is also different. We do not at this time understand
the origin
of this disagreement.

Consider the diagrams shown in figure 1. Since we are interested in a
long
distance effect, and correspondingly low momentum exchange, it is a
good
approximation to combine the effects of W and Z exchange
into four-fermi operators involving neutrinos and weakly charged
source
particles. The resulting operator can be Fierz transformed into the
following
form:
\begin{equation}
\O_4 = \frac{G_F}{ \sqrt{2} }
   \bigr[ \bar{\nu} \gamma_{\mu} ( 1 - \gamma_5 ) \nu \bigl]
   \bigr[ \bar{u} \gamma^{\mu} ( a - b \gamma_5 ) u \bigl],
\end{equation}
where a and b depend on the fermion u.

If u is an electron, both W and Z exchange contribute, yielding
$a = 2 sin^2 \theta_w + 1/2$ and $b = 1/2$. For a nucleon, only Z
exchange
contributes, yielding $a = -1/2, b = -g_A/2$ for the neutron and

$a= - 2 sin^2 \theta_w + 1/2, b = g_A/2$ for the
proton.  $g_A$ is the isotriplet axial vector form factor, with a
value

of order 1.25 .  W exchange
alone produces $a=b=1$ for electrons.

It remains to compute the diagram which results from two insertions
of the operator $\O_4$. The resulting S matrix element can be
directly
related to the interaction potential in the non-relativistic limit.
We will use momentum space Feynman rules.
However, working directly in coordinate space allows one to quickly
deduce
the $1/r^5$ form of the potential. In coordinate space massless
fermion
propagators behave like $1/x^3$, and one integrates over all time to
obtain
the potential. Thus we have \footnote{We thank S. Coleman for
pointing out
this result. It had been noted previously by J. Sucher in \cite{F}.}
\begin{equation}
\VN \sim G_F^2 \int dt~ x^{-6} \sim G_F^2 ~ x^{-5}.
\end{equation}

The momentum space amplitude for figure 1 is
\begin{equation}
T(q) = i G_F^2 \bigl[ \bar{u}(p_4) \gamma^{\mu} ( a - b \gamma_5 )
u(p_3) \bigr]
 \bigl[ \bar{u}(p_2) \gamma^{\nu} ( a' - b' \gamma_5 ) u(p_1) \bigr]
\Pi_{\mu \nu} (q),
\end{equation}
where $q = p_2 - p_1 = p_4 - p_3$ is the momentum transfer.
The polarization tensor $\Pi_{\mu \nu}(q)$ results from the fermion
loop.
The result is $1/2$ times the vacuum polarization tensor which
results
from the non-chiral loop, which can
be found in standard textbooks \cite{text} on QED. By current
conservation
$\Pi_{\mu \nu} (q) = A(q^2) ( q_{\mu} q_{\nu} -
g_{\mu \nu} q^2 ) $, where $A(q^2)$ is a dimensionless function of
$q^2$.
To obtain $\VN$ it is necessary to take the Fourier transform
\begin{equation}
\VN = i \int \frac {d^3 q} {(2 \pi)^3} e^{i \vec{q} \cdot \vec{r} } T
(q)|_{q^0 = 0}.
\end{equation}
The piece of the vacuum
polarization which leads to a long range interaction must depend on a
logarithm of $q^2$. This is because the Fourier transform of a
function
which is a polynomial in $q_{\mu}$ can always be written as
derivatives of a
delta function ( $ \partial_{\mu} \delta (r) $ ), which yield only
contact
interactions. The Fourier transform of a polynomial $\times$ log
function of
q can be shown to lead to $1/r^k$ type terms.

The relevant piece of $\Pi_{\mu \nu} (q)$ is readily extracted. This
yields
\begin{equation}
\Pi_{\mu \nu} (q) = - \frac{1}{24 \pi^2}
    ( q_{\mu} q_{\nu} - g_{\mu \nu} q^2 ) ln( q^2) + polynomial.
\end{equation}
Substituting this into the expression for $T$ yields several terms.
Consider
the terms which result from the contraction of $q_{\mu}$ with
$ \gamma^{\mu} ( a - b \gamma_5 ) $. The $q_{\mu} \gamma^{\mu}$ term
is zero
by the equations of motion, while the $q_{\mu} \gamma^{\mu} \gamma_5$
term
yields (in the non-relativistic approximation)
$\phi^{\dagger} \bigl( \sigma \cdot q \bigr) \phi$, where $\phi$ is a
two
component Pauli spinor. The $g_{\mu \nu}$ terms yield the
monopole-monopole term along with more complicated spin dependent
terms.

The Fourier transform we need to extract the potentials from $T$ is
given by
\begin{equation}
I_{ij} (r) ~\equiv~ \int \frac{ d^3 q }{ (2 \pi)^3 }~
    log q~ q_i q_j~ e^{i q \cdot r} \\
         ~=~ \frac{1}{4 \pi} \bigl[ 15 \frac{r_i r_j}{r^7}
             - 3 \frac{ \delta_{ij} }{ r^5 }
             \bigr].
\end{equation}
Using the above equation, we have the following result for $\VN$.
\begin{equation}
\VN = \frac{G_F^2}{8 \pi^3 r^5} \Bigl[ aa' - bb'
   \bigl(  1/2~ \sigma_1 \cdot \sigma_2 + 5/2 ~(\hat{r} \cdot
\sigma_1)
             (\hat{r} \cdot \sigma_2)  \bigr) + ...  \Bigr],
\end{equation}
where the ellipsis denote more complicated terms which are
proportional to
velocity and are suppressed in the non-relativistic limit, or which
are of
order $1/r^6$ or higher.
Note that the
monopole-monople interaction is repulsive between two electrons.
The sign of the interaction is changed if one of the particles is
replaced
by its antiparticle.

Let us briefly consider the phenomenological relevance of the above
interaction. The two neutrino force discussed here, which couples to
both
nucleons and electrons, is unscreened in normal
matter. However,
as mentioned previously, a $1/r^5$ potential is extremely weak.
The adjective `long range' is perhaps misleading as even large
volumes are
not capable of enhancing the effect of the force. For two nucleons
$\VN$
becomes comparable to the gravitational potential energy only at
separations
$r \sim 10^{-6} cm$. Detection of the two neutrino force would
therefore
require laboratory tests of gravity on distance scales far smaller
and at
sensitivities far greater than thus
far realized.  It seems highly unlikely that this effect will be
observed
in the forseeable future.

It is straightforward to see that the exchange of two massless
fermions
can at best lead to a $1/r^5$ potential for fermionic matter.
Consider a general four-fermi
interaction given by
\begin{equation}
\O_d = 1/M^{d+2}~ \bar{u} \Gamma_1 u~ \bar{\psi} \Gamma_2 \psi.
\end{equation}
Here the $\psi$'s are the massless fermions mediating the force,
$\Gamma_1,\Gamma_2$ are generic Lorentz tensors of total dimension d
(they may
contain derivatives), and M is some heavy mass scale.
By dimensional analysis the amplitude T
resulting from two insertions of $\O_d$ must have the form
\begin{equation}
1/M^{2d+4}~ A(q^2) ( q ...q )~
      \bar{u} \bar{\Gamma}_1 u ~\bar{u} \bar{\Gamma}_1 u,
\end{equation}
where the $q$'s are contracted with the $\bar{\Gamma}_1$ tensors,
which are
now dimensionless, and $A(q^2)$ is at most logarithmically dependent
on $q^2$.
Taking the Fourier transform of the above
amplitude results in a potential which falls as $1/r^{5+2d}$.

In the most general case of an arbitrary four-fermion vertex and a
${\it massive}$ fermion it is possible to generate an interaction
potential
of the form $V(r) \sim m^2 ~ e^{-2mr}/M^4 r^3$, where $M$ is the
large mass
scale suppressing the four-fermion interaction and m is the neutrino
mass.
Note that $m^2 e^{-2mr}$ has replaced $1/r^2$ in the massive case.
However,
it is easy to see that even for a fermion mass which is fine tuned to
some
inverse distance scale $r_*$, the strength of the massive interaction
cannot be significantly enhanced over the massless one.


\vskip .5 in
The authors would like to thank the Aspen Center for Physics, where
this
work was begun, for its kind
hospitality. While at the Aspen Center, the
authors benefitted from conversations with D. Caldwell,
S. Coleman, G. Gelmini, B. Kayser,
A. Kostelecky, P. Lanou, P. Rosen and L. Wolfenstein.
SDH acknowledges
support from the National Science Foundation under grant
NSF-PHY-87-14654,
the state of Texas under grant TNRLC-RGFY106, The Milton Fund of the
Harvard Medical School and from the Harvard Society
of Fellows. SDH would also like to thank the Physics Department at
Iowa
State University for its hospitality while this work was completed.
P. Sikivie acknowledges support from the DOE under DE-FG05-86ER40272.

\vskip 1 in


\end{document}